\documentclass[12pt]{article}
\usepackage[letterpaper,top=0.75in,bottom=0.75in,left=1in,right=1in,marginparwidth=0.5in]{geometry}

\usepackage{graphicx}
\usepackage{caption}
\usepackage{subcaption}
\usepackage{amsmath}
\usepackage{authblk}
\usepackage{float}

\usepackage[final]{pdfpages}

\begin{document}

\date{}
\title{Cross-Frequency Coupling Increases Memory Capacity in Oscillatory Neural Networks}

\author[1]{Connor Bybee}

\author[1]{Alexander Belsten}

\author[1, 2]{Friedrich T. Sommer}

\affil[1]{%
  {University of California at Berkeley}
  {Berkeley}
  {CA}
  {USA}
}

\affil[2]{%
  {Intel Labs}
  {Santa Clara}
  {CA}
  {USA}
}

\maketitle

\textbf{Abstract:} An open problem in neuroscience is to explain the functional role of oscillations in neural networks, contributing, for example, to perception, attention, and memory. Cross-frequency coupling (CFC) is associated with information integration across populations of neurons \cite{canolty2010functional}. Impaired CFC is linked to neurological disease \cite{zhang2016impaired}. It is unclear what role CFC has in information processing and brain functional connectivity. We construct a model of CFC which predicts a computational role for observed $\theta - \gamma$ oscillatory circuits in the hippocampus and cortex \cite{nandi2019inferring}. Our model predicts that the complex dynamics in recurrent and feedforward networks of coupled oscillators performs robust information storage and pattern retrieval.  Based on phasor associative memories (PAM) \cite{noest1987phasor}, we present a novel oscillator neural network (ONN) model \cite{hoppensteadt1998thalamo} that includes subharmonic injection locking (SHIL) \cite{Nishikawa2004} and which reproduces experimental observations of CFC. We show that the presence of CFC increases the memory capacity of a population of neurons connected by plastic synapses. CFC enables error-free pattern retrieval whereas pattern retrieval fails without CFC. In addition, the trade-offs between sparse connectivity, capacity, and information per connection are identified. The associative memory is based on a complex-valued neural network, or phasor neural network (PNN). We show that for values of $Q$ which are the same as the ratio of $\gamma$ to $\theta$ oscillations observed in the hippocampus and the cortex, the associative memory achieves greater capacity and information storage than previous models. The novel contributions of this work are providing a computational framework based on oscillator dynamics which predicts the functional role of neural oscillations and connecting concepts in neural network theory and dynamical system theory.

\begin{figure}[H]
    \centering
    \includegraphics[width=0.4\columnwidth]{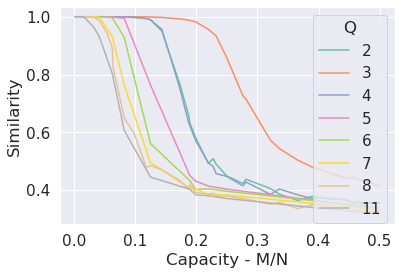}
    \includegraphics[width=0.4\textwidth]{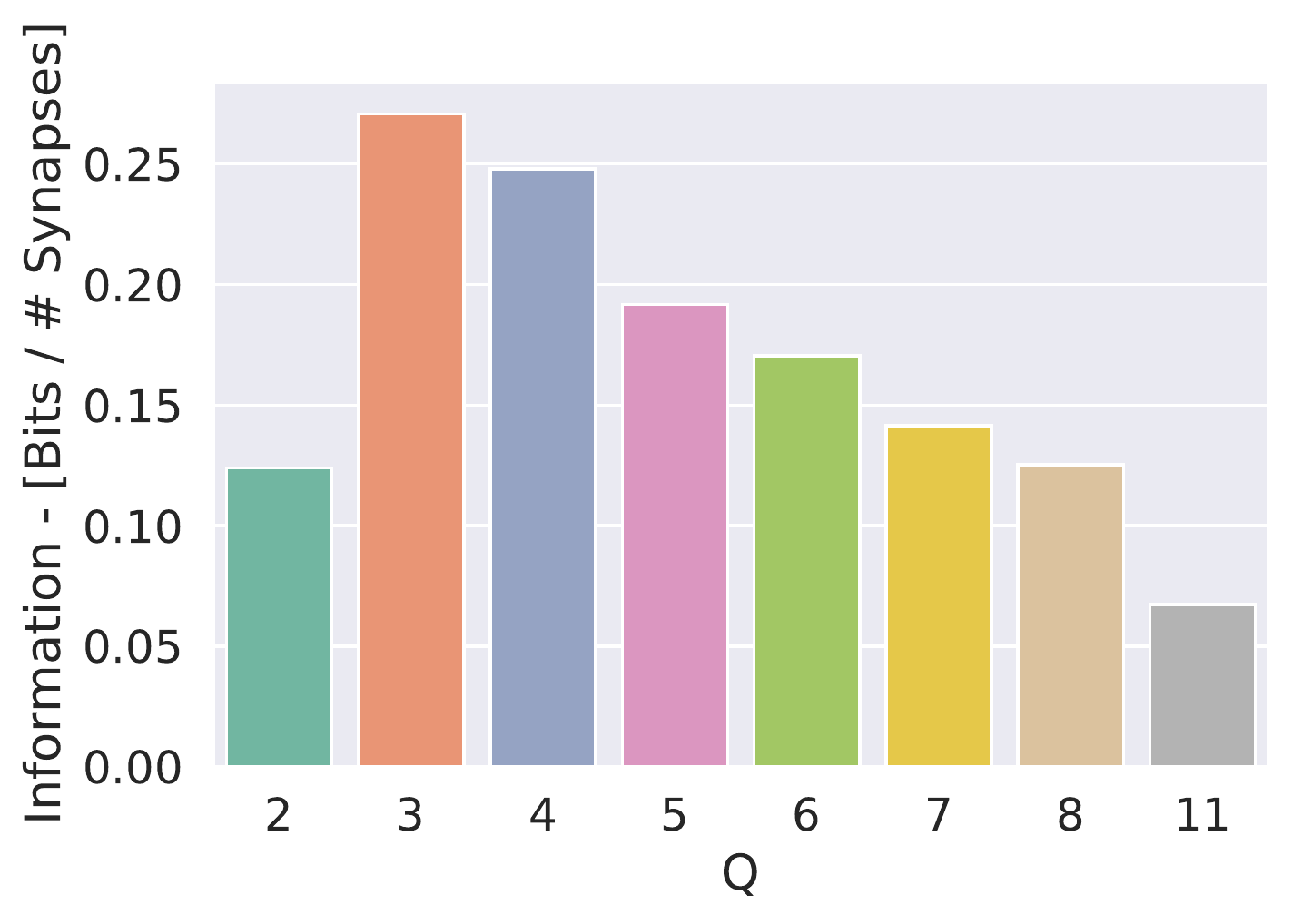}
    \includegraphics[width=0.4\columnwidth]{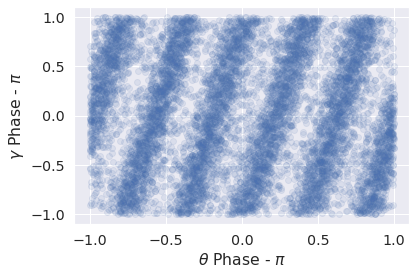}
    \includegraphics[width=0.4\columnwidth]{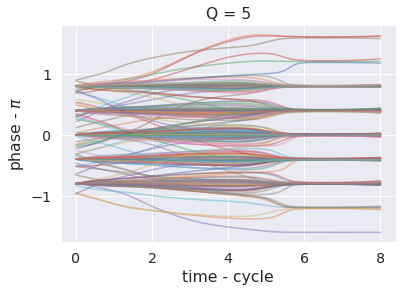}
    \caption{(Top) Results from simulation of $Q$-state PAMs with $N=1024$ units. The memory capacity is plotted against the pattern similarity upon retrieval (Top-Left) and synaptic information (Top-Right), measured in bits / synapse for error-free retrieval. Different modes of operation are possible depending on the value of $Q$. Greater values of $Q$ result in neural states with greater entropy, increasing information per pattern. Greater values of $Q$ tend to decreased memory capacity due to increased interference. $Q=3$ results in optimal storage. (Bottom-Left) A density plot of $\theta$-phase versus $\gamma$-phase demonstrates an example of CFC in a population of $\theta$ neurons driven by $\gamma$ neurons with additive white noise. (Bottom-Right) Simulation of an ONN with CFC. CFC is increased during the simulation. Without CFC, the network drifts from a stored pattern. After addition of CFC the network synchronizes and converges to a stored pattern.}
    \label{fig:phasor_capacity}
\end{figure}

\textbf{Description:} Identifying the mechanisms which integrate activity across neural circuits is needed in order to understand cognitive processes. CFC is the observation of correlations between multiple frequencies in neural oscillations. Our model proposes a computational role of CFC in ONNs which implement a PAM. Associative memories have been studied extensively as neural models of robust pattern storage and retrieval \cite{hopfield1982neural,Frady2019}. Storage capacity ($\frac{\text{M patterns}}{\text{N dimensions}}$) and synaptic information ($\frac{\text{bits}}{\text{synapse}}$) are key metrics determining the efficiency of an associative memory. A $Q$-state associative memory is a complex-valued PNN \cite{noest1988discrete} which stores a set of $M$  $N$-dimensional phasor patterns with discrete phases $\mathbf{\xi}^{k} = e^{\frac{i 2 \pi}{Q}  \mathbf{q}^{k}} \in \mathcal{C}^N, \ \forall k \in \{1,..,M\}$ where the phases have $Q$ discrete states $ q_{i}^{k} \in \{0,..,Q-1\} $. The energy, $E = -\frac{1}{2}\mathbf{z}^{* T}\mathbf{W} \mathbf{z}$, governs the system dynamics which move to stable, fixed-points in complex space, corresponding to limit-cycle attractors. Hebbian learning stores patterns in the weights using the complex outer product, $\mathbf{W} = \frac{1}{m}\sum_{k=1}^{m} \mathbf{\xi}^{k}\mathbf{\xi}^{k * T}$. Association is performed through recurrent dynamics using the update rule $ z_i(t+1) = f(\sum_{j} W_{ij}z_{j}(t))$. The activation function normalizes the amplitude and quantizes the phase to the nearest value of $q$, $f(u_i) = \text{exp}({i \frac{2\pi}{Q} \ \text{argmin}_{q} 
|\phi_{i}^{u} - \frac{2\pi q}{Q})| })$. Here, $u_i$ is the preactivation for unit $i$. The capacity and synaptic information are presented in figure \ref{fig:phasor_capacity}. PNNs can be mapped to models of physical systems, such as ONNs. Previous efforts to map PAMs to ONNs suffered from low capacity and poor pattern retrieval when the number of patterns exceeded 2. This results from the fixed-points of the dynamical system not being points corresponding to stored patterns. This issue can be corrected using SHIL, where the frequency of the harmonic injection signal is $Q$ times the frequency of memory oscillators, $Q = \omega_{2}/\omega_{1}$ . Our model is the first to implement $Q$-state PAMs in ONNs. PAMs implemented in ONNs with SHIL predict a functional role for CFC in the brain. The model can be adjusted to reproduce multiple types of CFC such as phase-amplitude coupling (PAC) and phase-phase coupling (PPC). As a simple example, consider two populations of oscillators, $\theta$ and $\gamma$, with different frequencies $\omega_{\theta}$ and $\omega_{\gamma}$, receptively. In the weakly-coupled limit, each oscillator can be represented as a phase, $\phi_i$, oscillating at frequency $\omega_i$. The phase of the theta oscillators change based on recurrent interactions and external drive by $\gamma$ oscillators. For the simple model, the system of ordinary differential equations (ODE) is $ \frac{\partial \phi_i}{\partial t} = -\epsilon \sum_{j \neq i} R_{ij} \sin(\phi_i - \phi_j - \Phi_{ij}) - h \sin(\omega \phi_i) $. Models with the appropriate parameters allow for complicated dynamics such as PAC, PPC, and out-of-phase but synchronized behavior.

This work is a step towards explaining the computational abilities of complex ONNs and the mechanisms of oscillatory neural circuits. PNNs connect low-level representations, such as grid-cells, place-cells, and phase-precession, with higher-level cognitive architectures for symbolic computation with high-dimensional vectors, such as Vector Symbolic Architectures, and Hyperdimensional Computing. To our knowledge, this is the first work to validate the capacity analysis of $Q$-state PAM networks \cite{Cook1989} through simulation and provide a model of how to implement such a memory on a physical system. Future work shall investigate spiking neuron models, the representational capabilities of PNNs, and mapping these to predictive models for neuroscience. 

\bibliographystyle{ieeetr}
\bibliography{references.bib}


\end{document}